\documentclass[a4paper,preprint]{aastex}
\usepackage{graphicx}
\begin{document}

\title{The Synchrotron Spectrum of Fast Cooling Electrons Revisited}

\author{Jonathan Granot\altaffilmark{1}, Tsvi Piran\altaffilmark{1}
  and Re'em Sari\altaffilmark{2}}

\altaffiltext{1}{Racah Institute, Hebrew University, Jerusalem 91904, Israel}
\altaffiltext{2}{Theoretical Astrophysics, California Institute of Technology, 
Pasadena, CA 91125, USA}

\begin{abstract}
  We discuss the spectrum arising from synchrotron emission by fast
  cooling (FC) electrons, when fresh electrons are continually
  accelerated by a strong blast wave, into a power law distribution of
  energies. The FC spectrum was so far described by four power law
  segments divided by three break frequencies $\nu _{sa}<\nu _{c}<\nu
  _{m}$. This is valid for a homogeneous electron distribution.
  However, hot electrons are located right after the shock, while most
  electrons are farther down stream and have cooled. This spatial
  distribution changes the optically thick part of the spectrum,
  introducing a new break frequency, $\nu_{ac}<\nu_{sa}$, and a new
  spectral slope, $F_{\nu }\propto \nu^{11/8}$ for $\nu_{ac}<\nu
  <\nu_{sa}$. The familiar $F_{\nu }\propto \nu ^{2}$ holds only for
  $\nu<\nu_{ac}$. This ordering of the break frequencies is relevant
  for typical gamma-ray burst (GRB) afterglows in an ISM environment.
  Other possibilities arise for internal shocks or afterglows in dense
  circumstellar winds. We discuss possible implications of this
  spectrum for GRBs and their afterglows, in the context of the
  internal-external shock model. Observations of
  $F_{\nu}\propto\nu^{11/8}$ would enable us to probe scales much
  smaller than the typical size of the system, and constrain the
  amount of turbulent mixing behind the shock.
\end{abstract}


\authoremail{\ jgranot@nikki.fiz.huji.ac.il
tsvi@nikki.fiz.huji.ac.il sari@tapir.caltech.edu}

\section{Introduction.}

The spectrum of Gamma-Ray Bursts (GRBs) and their afterglows is well
described by synchrotron and inverse Compton emission. It is better
studied during the afterglow stage, where we have broad band
observations. The observed behavior is in good agreement with the
theory. Within the fireball model, both the GRB and its afterglow are
due to the deceleration of a relativistic flow. The radiation is
emitted by relativistic electrons within the shocked
regions. According to the internal-external shock scenario, it has
been shown \citep{FMN,SP97} that in order to obtain a reasonable
efficiency, the GRB itself must arise from internal shocks (ISs)
within the flow, while the afterglow is due to the external shock (ES)
produced as the flow is decelerated upon collision with the ambient
medium. In the simplest version of the fireball model, a spherical
blast wave expands into a cold and homogeneous ambient medium (Waxman
1997; M\'{e}sz\'{a}ros, P. \& Rees, M.  1997; Katz \& Piran 1997;
Sari, Piran \& Narayan 1998, hereafter SPN).
An important variation is a density profile $\rho(r)\propto r^{-2}$,
suitable for a massive star progenitor which is surrounded by its
pre-explosion wind.

In this letter we consider fast cooling (FC), where the electrons cool
due to radiation losses on a time scale much shorter than the
dynamical time of the system\footnote{$t_{dyn}$ is the time for
considerable expansion. Adiabatic cooling is therefore negligible
compared to radiative cooling.}, $t_{dyn}$. Both the highly variable
temporal structure of most bursts and the requirement of a reasonable
radiative efficiency, suggest FC during the GRB itself 
\citep{SNP96}. During the afterglow, FC lasts $\sim 1$ hour after the
burst for an ISM surrounding (SPN; Granot, Piran \& Sari 1999a) and
$\sim 1$ day in a dense circumstellar wind environment \citep{C&L}.
We assume that the electrons (initially) and the magnetic field
(always) hold fractions $\epsilon_e$ and $\epsilon_B$ of the internal
energy, respectively. We consider synchrotron emission of relativistic
electrons which are accelerated by a strong blast wave into a power
law energy distribution:
$N(\gamma)\propto\gamma^{-p}$ for
$\gamma\ge\gamma_m = (p-2)\epsilon_e e' {\big /} (p-1)n' m_e c^2\cong
\epsilon_e e'{\big /} 3 n' m_e c^2$, 
where $n'$ and $e'$ are the number density and internal energy density
in the local frame and we have used the standard value, $p=2.5$.

After being accelerated by the shock, the electrons cool due to
synchrotron radiation losses. 
An electron with a critical Lorentz
factor, $\gamma_c$, cools on the dynamical time, $t_{dyn}$
\begin{equation}
\gamma_c=6\pi m_e c {\big /} \sigma_T\Gamma B'^{2}t_{dyn}=
3 m_e c {\big /} 4 \sigma_T\Gamma \epsilon_B e' t_{dyn}\ ,
\label{gamma_c}
\end{equation}
where $\Gamma$ is the bulk Lorentz factor, $\sigma_T$ is the Thomson
cross section and $B'$ is the magnetic field.  The Lorentz factors,
$\gamma_m$ and $\gamma_c$, correspond to the frequencies: $\nu_m$ and
$\nu_c$, respectively, using $\nu_{syn}(\gamma)= 3 q_e B'\gamma^2
\Gamma {\big /} 16 m_e c$. FC implies that $\gamma_c \ll \gamma_m$ and
therefore $\nu_c \ll \nu_m$.

The FC spectrum had so far been investigated only for
$\nu_{sa}<\nu_c<\nu_m$, using a homogeneous distribution of electrons
(SPN, Sari \& Piran 1999), where $\nu_{sa}$ is the self absorption
frequency. Under these assumptions
the spectrum consists of four power law segments:
$F_{\nu}\propto\nu^{2}, \nu^{1/3}, \nu^{-1/2}$ and $\nu^{-p/2}$, from
low to high frequencies. The spectral slope above $\nu_m$ is related
to the electron injection distribution: the number of electrons with
Lorentz factors $\sim\gamma$ is $\propto\gamma^{1-p}$ and their energy
$\propto\gamma^{2-p}$. As these electrons cool, they deposit most of
their energy into a frequency range
$\sim\nu_{syn}(\gamma)\propto\gamma^2$ and therefore
$F_{\nu}\propto\gamma^{-p}\propto\nu^{-p/2}$. At $\nu _c<\nu<\nu_m$
all the electrons in the system contribute, as they all cool on the
dynamical time, $t_{dyn}$. Since the energy of an electron
$\propto\gamma$, and its typical frequency $\propto\gamma^2$ the flux
per unit frequency is $\propto\gamma^{-1}\propto \nu^{-1/2}$.  The
synchrotron low frequency tail of the cooled electrons ($\propto
\nu^{1/3}$) appears at $\nu_{sa}<\nu<\nu_c$.  Below $\nu_{sa}$, the
the system is optically thick to self absorption and we see the
Rayleigh-Jeans portion of the black body spectrum:
\begin{equation}
\label{RJ}
F_{\nu}\propto\nu^2\gamma_{typ}(\nu) \ , 
\end{equation}
where $\gamma_{typ}(\nu)$ is the typical Lorentz factor of the
electrons emitting at the observed frequency $\nu$. Assuming
$\gamma_{typ}(\nu)=\gamma_c\propto\nu^{0}$, one obtains
$F_{\nu}\propto\nu^2$.


We derive the FC spectrum of an inhomogeneous electron temperature
distribution  in \S 2 . We find a
new self absorption regime where $F_{\nu}\propto\nu^{11/8}$. In \S 3
we calculate  the  break frequencies and flux
densities for ESs (afterglows) with a spherical adiabatic evolution,
both for a homogeneous external medium and for a stellar wind
environment.  ISs are treated in \S 4.  In section \S 5, we show that
the early radio afterglow observations may be affected by the new
spectra. We find that synchrotron self absorption is unlikely to
produce
the steep slopes observed in some bursts in the
$1-10$ keV range. We also discuss the possibility of using the
new spectra to probe very small scales behind the shock.


\section{Fast cooling spectrum}

\label{FCS}

The shape of the FC spectrum is determined by the relative ordering of
$\nu_{sa}$ with respect to $\nu_c<\nu_m$.  There are three possible
cases.  We begin with $\nu_{sa}<\nu_c<\nu_m$, case 1 hereafter. This
is the ``canonical'' situation, which arises for a reasonable choice
of parameters for afterglows in an ISM environment.  
The optically thin part of the spectrum ($\nu>\nu_{sa}$) of an
inhomogeneous electron distribution is similar to the homogeneous one.
All the photons emitted in this regime escape the system, rendering
the location of the emitting electrons unimportant. In the optically
thick regime ($\nu<\nu_{sa}$), most of the escaping photons are
emitted at an optical depth $\tau_{\nu} \sim 1$, and
$\gamma_{typ}(\nu)$ must be evaluated at the place where
$\tau_{\nu}=1$.

In an ongoing shock there is a continuous supply of newly accelerated
electrons. These electrons are injected right behind the shock with
Lorentz factors $\gamma\geq\gamma_m$, and then begin to cool due to
radiation losses. In the relativistic shock frame, the shocked fluid
moves backwards at a speed of $c/3$: $l'=ct'/3$, where $l'$ is the
distance of a fluid element behind the shock and $t'$ is the time
since it passed the shock. Just behind the shock there is a thin layer
where the electrons have not had sufficient time to cool
significantly. Behind this thin layer there is a much wider layer of
cooled electrons. All these electrons have approximately the same
Lorentz factor: $\gamma(t')=6\pi m_e c/\sigma_T B^{\prime}{}^{2}t'$,
or equivalently, $\gamma(l')=2\pi m_e c^2/\sigma_T
B^{\prime}{}^{2}l'$.  Electrons that were injected early on and have
cooled down to $\gamma_c$ are located at the back of the shell, at a
distance of $\Delta'=ct_{dyn}'/3=c\Gamma t_{dyn}/3=\Gamma\Delta$
behind the shock
(i.e. $\gamma_c=\gamma(\Delta')=\gamma(t'_{dyn})$). We define the
boundary between the two layers, $l'_{0}$, as the place where an
electron with an initial Lorentz factor $\gamma_m$ cools down to
$\gamma_m/2$
\begin{equation}
\label{l0}
l'_{0}=2\pi m_e c^2 {\big /} \sigma_T B^{\prime}{}^{2}\gamma_m
=3 m_e^2 c^4 n^{\prime}{\big/}
4\sigma_T\epsilon_e\epsilon_B e^{\prime}{}^{2} \ .
\end{equation}
The un-cooled layer is indeed very thin, as
$l'_{0}/\Delta'=\gamma_c/\gamma_m\ll 1$.  We define $l'_1(\nu)$ by
$\tau_{\nu}(l'_1) = 1$.  The optically thin emission from $l'<l'_1$
equals the optically thick emission:
\begin{equation}
n'l'_{1}\Gamma^2 P_{\nu,max}\left[\nu/\nu_{syn}(\gamma(l'_1))\right]^{1/3}
{\big/}4\pi=\left(2\nu^2/ c^2\right)\Gamma\gamma(l'_1) m_e c^2\ , \label{tau=1}
\end{equation}
where $P_{\nu,max}\approx
P_{syn}(\gamma)/\nu_{syn}(\gamma)\propto\gamma^0$ and
$P_{syn}(\gamma)=\Gamma^2 \sigma_T c \gamma^2 B^{\prime}{}^{2}{\big
/}6\pi$ are the peak spectral power and total synchrotron power of an
electron, respectively. Since $\nu _{syn}(\gamma)\propto\gamma^2$, and
within the cooled layer, $\gamma=\gamma(l')\propto1/l'$, eq.
(\ref{tau=1}) implies $\gamma_{typ}=\gamma(l'_1)\propto\nu^{-5/8}$.
We now use eq. (\ref{RJ}) and obtain that $F_{\nu }\propto \nu
^{11/8}$.  This new spectral regime, is a black body spectrum,
modified by the fact that the effective temperature ($\gamma_{typ}$)
varies with frequency.

At sufficiently low frequencies, $l'_1(\nu)<l'_{0}$, implying
$\gamma_{typ}=\gamma_m\propto\nu^0$ and $F_{\nu}\propto\nu^2$. The
transition from absorption by the cooled electrons ($F_{\nu}\propto
\nu ^{11/8}$) to absorption by un-cooled electrons ($F_{\nu }\propto
\nu ^{2}$) is at $\nu_{ac}$, which satisfies $l'_1(\nu_{ac})=l'_{0}$. The
resulting spectrum is shown in the upper frame of Fig. \ref{Fig1}.
$\nu_{ac}$ may be obtained from eq.  (\ref{tau=1}) by substituting
$\gamma(l'_1)=\gamma_m$
and $l'_1=l'_{0}$ from eq. (\ref{l0}).  Substituting
$\gamma(l'_1)=\gamma_c$ from eq.  (\ref{gamma_c}) and
$l'_1=\Delta'=\Gamma\Delta$ into eq.  (\ref{tau=1}), gives us
$\nu_{sa}^{(1)}$. The superscript, $^{(i)}$, labels the specific case
under consideration. We obtain
\begin{eqnarray}
\nu _{ac} &=& \left(6m_{e}^{12}c^{29}\gamma^{5}n^{\prime}{}^{11}{\big /}
\pi ^{5}\epsilon _{B}^{2}\epsilon _{e}^{8}q_{e}^{4}e^{\prime }{}^{10}
\right)^{1/5}\ , \label{nu_ac} \nonumber 
\\
\nu _{sa}^{(1)} &=& \left(8/3\pi\right)\left[4\sigma_T^8\epsilon_B^6
\Gamma^{13}\Delta^8 n^{\prime}{}^{3}e^{\prime}{}^{6}{\big /} 
9 m_e^4 c^3 q_e^4\right]^{1/5}\ . \label{nu_sa1} 
\end{eqnarray}
The ratio
$\nu^{(1)}_{sa}/\nu_{ac}=(\gamma_m/\gamma_c)^{8/5}=(\nu_m/\nu_c)^{4/5}$
depends on the cooling rate. The maximal flux density occurs at
$\nu_{c}=\nu _{syn}(\gamma_c)$ and is given by
$F^{(1)}_{\nu,max}=N_{e}P_{\nu,max}/4\pi D^{2}$ (SPN), where $N_{e}$
is the number of emitting electrons and $D$ is the distance to the
observer, while $\nu_m=\nu_{syn}(\gamma_m)$
\begin{eqnarray}
\nu_c &=& 27\sqrt{\pi}m_e c q_e\ {\big /}\ 64\sqrt{2}\sigma_T^2\Gamma 
\epsilon_B^{3/2}t_{dyn}^2 e^{\prime}{}^{3/2}\ ,  \label{nu_c}
\nonumber \\
\nu_m &=& \sqrt{\pi} q_e \Gamma \epsilon_B^{1/2} \epsilon_e^2 
e^{\prime}{}^{5/2} \ {\big /}\ 12 \sqrt{2} m_e^3 c^5 
n^{\prime}{}^2\ , \label{nu_m} \\
F_{\nu ,max}^{(1)} &=& \ 4\sqrt{2}\sigma_T m_e c^2
N_{e}\gamma \epsilon _{B}^{1/2}e^{\prime}{}^{1/2}
\ {\big /}\ 9\pi^{3/2} q_e D^2\ \nonumber .  
\end{eqnarray}



For  $\nu _{c}<\nu _{sa}<\nu _{m}$ (case 2)  the cooling
frequency, $\nu_{c}$, becomes unimportant, as it lies in the optically
thick regime. Now there are only three transition frequencies. 
$\nu_{ac}$
and $\nu _{m}$ are similar to case 1. The peak flux, 
$F_{\nu ,max}^{(2)}$ is reached at $\nu_{sa}^{(2)}$: 
\begin{eqnarray}
\nu _{sa}^{(2)} &=& (\nu _{sa}^{(1)})^{5/9}\nu _{c}^{4/9}\ ,
\nonumber \\
F_{\nu ,max}^{(2)} &=& F_{\nu ,max}^{(1)}
\left(\nu _{c}/\nu_{sa}^{(1)}\right)^{5/18}\  \ .
\end{eqnarray}



If $\nu _{c}<\nu _{m}<\nu _{sa}$ (case 3) then $l'_1(\nu)\ll l'_{0}$
for $\nu<\nu_{sa}^{(3)}$. Now, both $\nu _{ac}$ and $\nu _{c}$ are
irrelevant, as the inner parts, where these frequencies are important,
are not visible. We can use the initial electron distribution to
estimate $\gamma_{typ}$: $\gamma_{typ}=\gamma_{m}\propto\nu^{0}$ at
$\nu<\nu_{m}$, implying $F_{\nu}\propto \nu ^{2}$. At $\nu_{m}<\nu
<\nu _{sa}$ the emission is dominated by electrons with
$\nu_{syn}(\gamma)\sim \nu$, implying $\gamma_{typ}\propto\nu^{1/2}$
and $F_{\nu}\propto\nu^{5/2}$.  $F_{\nu,max}^{(3)}$ is reached at $\nu
_{sa}^{(3)}$:
\begin{eqnarray}
\nu _{sa}^{(3)} &=& \left[ {(\nu _{sa}^{(1)})^{10/3}\nu
_{c}^{8/3}\nu _{m}^{p-1}}\right]^{1/(p+5)}\ ,  \nonumber \\
F_{\nu ,max}^{(3)} &=& F_{\nu ,max}^{(1)}\left[{(\nu _{sa}^{(1)})^{-p/3}
\nu _{c}^{(3-p)/6}\nu _{m}^{(p-1)/2}}\right]
^{5/(p+5)}\ .  \label{case3}
\end{eqnarray}


\section{Application to External Shocks and the Afterglow}

Consider now the FC spectrum of an ES which is formed when a
relativistic flow decelerates as it sweeps the ambient medium. This is
the leading scenario GRB afterglow.  We consider an adiabatic
spherical outflow running into a cold ambient medium with a density
profile $\rho(r)\propto r^{-\alpha}$, for either $\alpha=0$
(homogeneous ISM) or $\alpha=2$ (stellar wind environment).  FC lasts
for the first hour or so in a typical ISM surrounding, and for about a
day in a stellar wind of a massive progenitor. The proper number
density and internal energy density behind the shock are given by the
shock jump conditions: $n^{\prime}=4\Gamma n$ and
$e^{\prime}=4\Gamma^{2}n m_{p}c^{2}$, where $n$ is the proper number
density before the shock and $m_p$ is the mass of a proton. We also
use $\Delta =R/12\Gamma^2$, $R=4\Gamma^2 ct$ and $t_{dyn}=t$, where
$t$ is the observed time.

For a homogeneous environment $N_{e}=4\pi n R^{3}/3$. Using $\Gamma
\propto t^{-3/8}$ (e.g. SPN) we obtain
\begin{eqnarray}
\nu _{ac} &=&1.7\times 10^{9}\ {\rm Hz}\ \epsilon _{B,0.1}^{-2/5}\epsilon
_{e,0.1}^{-8/5}E_{52}^{-1/10}n_{1}^{3/10}t_{2}^{3/10}\ ,  \nonumber
\label{ES_ISM} \\
\nu _{sa}^{(1)} &=&1.8\times 10^{10}\ {\rm Hz}\ \epsilon
_{B,0.1}^{6/5}E_{52}^{7/10}n_{1}^{11/10}t_{2}^{-1/2}\ ,  \nonumber \\
\nu _{c} &=&2.9\times 10^{15}\ {\rm Hz}\ \epsilon
_{B,0.1}^{-3/2}E_{52}^{-1/2}n_{1}^{-1}t_{2}^{-1/2}\ ,  \nonumber \\
\nu _{m} &=&5.5\times 10^{16}\ {\rm Hz}\ \epsilon _{B,0.1}^{1/2}\epsilon
_{e,0.1}^{2}E_{52}^{1/2}t_{2}^{-3/2}\ ,  \nonumber \\
F_{\nu ,max}^{(1)} &=&30\ {\rm mJy}\ D_{28}^{-2}\epsilon
_{B,0.1}^{1/2}E_{52}n_{1}^{1/2}\ ,
\end{eqnarray}
where $n_1=n/1 {\ \rm cm^{-3}}$, $E_{52}=E/10^{52}{\ \rm ergs}$,
$D_{28}=D/10^{28}{\ \rm cm}$, $\epsilon _{B,0.1}=\epsilon _{B}/0.1$,
$\epsilon _{e,0.1}=\epsilon _{e}/0.1$ and $t_{2}=t/100 {\ \rm sec}$.
For typical parameters, only the case 1 spectrum  
is expected. After $\sim 1$ hour, slow
cooling (SC) sets in, and the spectrum is given in GPS.

For a circumstellar wind environment, $n=r^{-2}A/m_p$ and $N_e=4\pi A
R/m_p$.  Using $\Gamma\propto t^{-1/4}$ and $A=5\times 10^{11}
A_{\star}{\ \rm gr\ cm^{-1}}$ as in Chevalier \& Li (1999) we obtain
\begin{eqnarray}
\nu _{ac} &=&3.6\times 10^{10}\ {\rm Hz}\ A_{\star}^{3/5}
\epsilon _{B,0.1}^{-2/5}\epsilon_{e,0.1}^{-8/5}E_{52}^{-2/5}\ ,  
\nonumber \label{ES_wind} \\
\nu _{sa}^{(1)} &=&8.0\times 10^{11}\ {\rm Hz}A_{\star}^{11/5}
\epsilon_{B,0.1}^{6/5}E_{52}^{-2/5}t_{hr}^{-8/5}\ ,  \nonumber \\
\nu _{c} &=&2.5\times 10^{12}\ {\rm Hz}\ A_{\star}^{-2}
\epsilon_{B,0.1}^{-3/2}E_{52}^{1/2}t_{hr}^{1/2}\ ,  \nonumber \\
\nu _{m} &=&1.2\times 10^{14}\ {\rm Hz}\ \epsilon _{B,0.1}^{1/2}\epsilon
_{e,0.1}^{2}E_{52}^{1/2}t_{hr}^{-3/2}\ ,  \nonumber \\
F_{\nu ,max}^{(1)} &=&0.39\ {\rm Jy}\ D_{28}^{-2}A_{\star}
\epsilon_{B,0.1}^{1/2}E_{52}^{1/2}t_{hr}^{-1/2}\ ,\nonumber \\
\nu _{sa}^{(2)} &=&1.3\times 10^{12}\ {\rm Hz}A_{\star}^{1/3}
t_{hr}^{-2/3}\ ,  \nonumber \\
F_{\nu ,max}^{(2)} &=&0.54\ {\rm Jy}\ D_{28}^{-2}A_{\star}^{-1/6}
\epsilon_{B,0.1}^{-1/4}E_{52}^{3/4}t_{hr}^{1/12}\ ,
\end{eqnarray}
where $t_{hr}=t/1\ {\rm hour}$. For typical parameters, the spectrum
is of case 2  for $\sim 1-2$
hours after the burst. Then it turns to  case 1  
until $\sim 1$ day, when there is a transition to
SC. The SC spectrum is given in Chevalier \& Li (1999).

\section{Application to Internal Shocks and the GRB}

ISs are believed to produce the GRBs themselves. The temporal
variability of the bursts is attributed to emission from many
different collisions between shells within the flow. The number of
peaks in a burst, $N$, roughly corresponds to the number of such
shells. Different shells typically collide before their initial width,
$\Delta_{i}$, has expanded significantly. Assuming that the typical
initial separation between shells is $\sim\Delta_{i}$, $\Delta_{i}
=cT_{90}/2N$ in average, where $T_{90}$ is the duration of the burst.
The average energy of a shell is $E_{sh}\approx E/N$, where $E$ is the
total energy of the relativistic flow. The emission in the optically
thick regime comes from the shocked fluid of the outer and slower
shells. We denote the initial Lorentz factor of this shell by
$\Gamma_{i}$, and its Lorentz factor after the passage of the shock by
$\Gamma$. The average thermal Lorentz factor of the protons in this
region equals the relative bulk Lorentz factor of the shocked and
un-shocked portions of the outer shell, $\Gamma
_{r}=\Gamma/2\Gamma_{i}$, which is typically of order unity.
Therefore $e^{\prime}=\Gamma_{r}n^{\prime}m_{p}c^{2}$. One can
estimate the number density of the pre-shocked fluid,
$n_{i}^{\prime}$, by the number of electrons in the shell,
$N_{e}=E_{sh}/\Gamma_{i}m_{p}c^{2}$, divided by its volume:
$n_{i}^{\prime}= N_{e}/4\pi R^{2}\Delta_{i}\Gamma_{i}$. The number
density of the shocked fluid, which is the one relevant for our
calculations, is $n^{\prime}=4\Gamma_r n_{i}^{\prime}$. The width of
the front shell in the observer frame decreases after it is shocked:
$\Delta=\Delta_{i}/8\Gamma_r^{2}$. In this section we use $R\sim
2\Gamma_{i}^{2}\Delta_{i}\sim 4\Gamma^2 \Delta $, which is the typical
radius for collision between shells, and $t_{dyn}\sim 3\Delta /c$.
Thus, we obtain
\begin{eqnarray}
\nu _{ac} &=&8.7\times 10^{14}\ {\rm Hz}\ \Gamma _{r,3}^{-3/5}\epsilon
_{B,0.1}^{-2/5}\epsilon _{e,0.1}^{-8/5}E_{52}^{1/5}\Gamma
_{3}^{-1/5}N_{2}^{2/5}T_{1}^{-3/5}\ , 
\nonumber \\
\nu _{sa}^{(1)} &=&7.7\times 10^{19}\ {\rm Hz}\ \Gamma _{r,3}^{53/5}\epsilon
_{B,0.1}^{6/5}E_{52}^{9/5}\Gamma _{3}^{-41/5}N_{2}^{2}T_{1}^{-19/5}\ , 
\nonumber  \label{IS} \\
\nu _{c} &=&3.6\times 10^{12}\ {\rm Hz}\ \Gamma _{r,3}^{-8}\epsilon
_{B,0.1}^{-3/2}E_{52}^{-3/2}\Gamma _{3}^{8}N_{2}^{-1}T_{1}^{5/2}\ , 
\nonumber \\
\nu_{m} &=&5.6\times 10^{18}\ {\rm Hz}\ \Gamma _{r,3}^{6}\epsilon
_{B,0.1}^{1/2}\epsilon _{e,0.1}^{2}E_{52}^{1/2}\Gamma
_{3}^{-2}N_{2}T_{1}^{-3/2}\ ,  
\nonumber \\
F_{\nu ,max}^{(1)} &=&509\ {\rm Jy}\ \Gamma _{r,3}^{5}\epsilon
_{B,0.1}^{1/2}E_{52}^{3/2}\Gamma _{3}^{-3}D_{28}^{-2}T_{1}^{-3/2}\ ,
\nonumber \\
\nu _{sa}^{(2)} &=&4.3\times 10^{16}\ {\rm Hz}\ \Gamma_{r,3}^{7/3}
E_{52}^{1/3}\Gamma_{3}^{-1}N_{2}^{2/3}T_{1}^{-1}\ ,
\end{eqnarray}
where $N_{2}=N/100$, $T_{1}=T_{90}/10\ {\rm sec}$, $\Gamma
_{r,3}=\Gamma_r/3$ and $\Gamma _{3}=\Gamma/10^{3}$.

\section{Discussion}

We have calculated the synchrotron spectrum of fast cooling (FC)
electrons.  We find three possible spectra, depending on the relative ordering
$\nu_{sa}$ with respect to $\nu_c<\nu_m$. Two of these spectra contain
a new self absorption regime where $F_{\nu}\propto\nu^{11/8}$.

\begin{figure}
\centering
\noindent
\includegraphics[width=13cm]{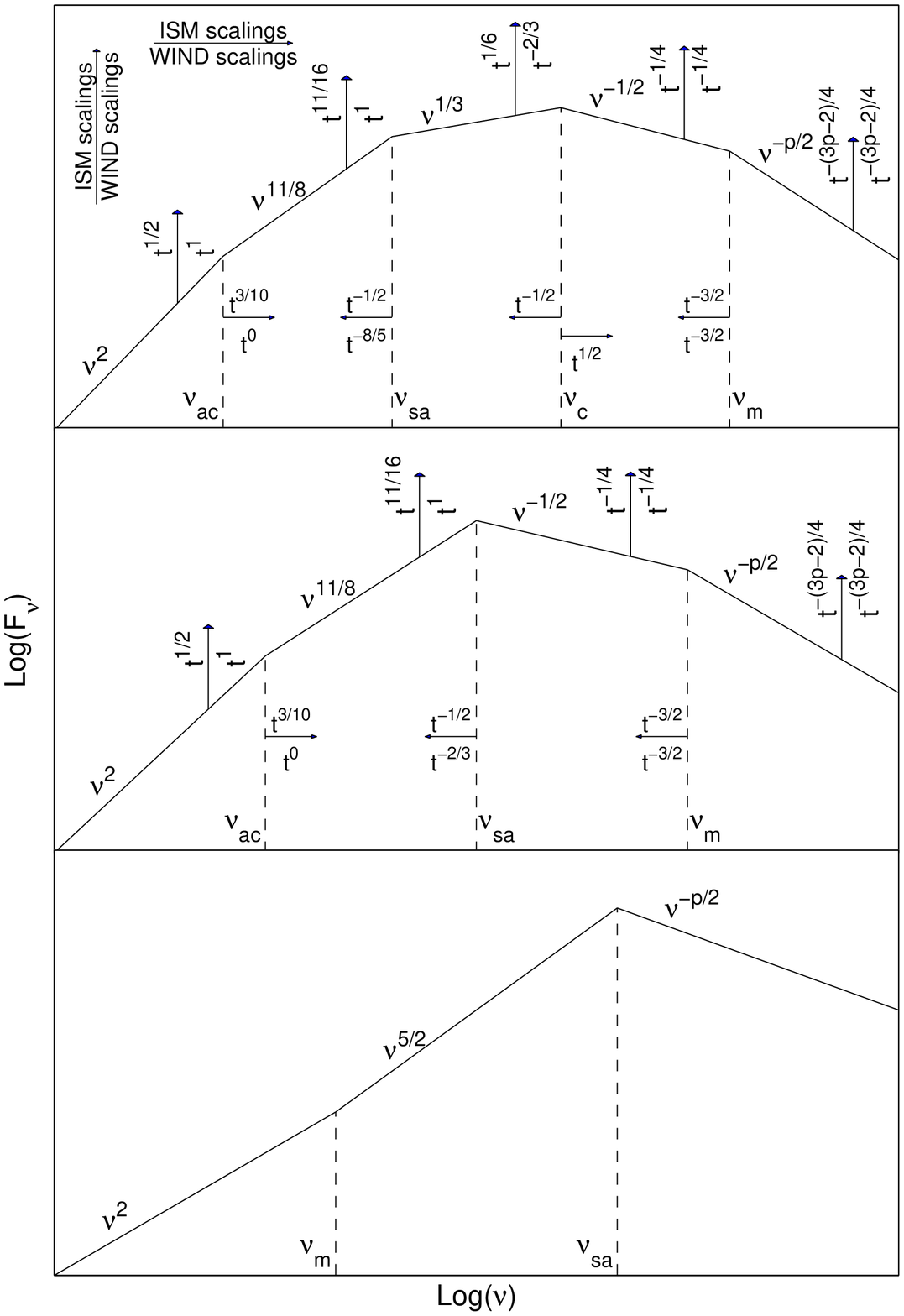}
\caption{\label{Fig1}Fast cooling (FC) synchrotron spectra from a shock 
  injected power law electron distribution. The shape of the spectrum
  is determined by the ordering of the self absorption frequency,
  $\nu_{sa}$, with respect to $\nu_c<\nu_m$. There are three possible
  shapes for the spectrum, corresponding to $\nu_{sa}<\nu_c$,
  $\nu_c<\nu_{sa}<\nu_m$ and $\nu_{sa}>\nu_m$, from top to bottom.
  Scalings for an afterglow, both in an ISM and wind environments, are
  given only in the first two frames, since for typical parameters
  $\nu_{sa}<\nu_m$ during the FC phase of the afterglow.}
\end{figure}

During the initial fast cooling stage of the afterglow, the system is
typically optically thick in the radio and optically thin in the
optical and X-ray, for both ISM and stellar wind environments. We
therefore expect the new feature, $F_{\nu }\propto \nu ^{11/8}$, to be
observable only in the radio band, during the afterglow. For both
environments, $\nu_{ac}$ and $\nu_{sa}$ move closer together (see Fig.
\ref{Fig1}) until they merge at the transition to slow cooling (SC).
Afterwards, there is only one self absorption break, at
$\nu_{sa}^{(SC)}$.

For an ISM surrounding, $\nu_{sa}^{(SC)}\propto t^0$. From current
late time radio observations, we know that typically,
$\nu_{sa}^{(SC)}\sim$ a few GHz \citep{Taylor,W&G,GPS2}. Therefore, the
whole VLA band, $1.4-15$ GHz, should initially be in the range where
$F_{\nu}\propto\nu^{11/8}$.  Sufficiently early radio observations,
which could confirm this new spectral slope, may become available in
the upcoming HETE era.

In a considerable fraction of bursts, there is evidence for a spectral
slope $>1/3$ (photon number slope $>-2/3$), in the $1-10$ keV range
\citep{Preece98,Crider,Strohmayer}. Such spectral slopes are not
possible for optically thin synchrotron emission \citep{Katz94}.
They could be explained by self absorbed synchrotron emission if
$\nu_{sa}$ reaches the X-ray band. The best prospects for this to
occur are with the spectrum of the second type.  However, we have to
check whether the physical parameters for which $\nu_{sa}$ is so high
are reasonable. Several constraints must be satisfied: (i) ISs must
occur at smaller radii than ESs, (ii) efficient emission requires FC,
and (iii) the system must be optically thin to Thomson scattering and
pair production.  The most severe constraint in the way of getting the
$\nu_{sa}$ into the X-ray band arises from (iii).  It is possible only
for rather extreme parameters: $\Gamma\sim 10^4$ and $\Gamma_r\sim
50$.  With such parameters, $\nu F_{\nu}$ would peak at $h\nu_m\sim
1-100{\ \rm GeV}$, unless $\epsilon_e\lesssim 10^{-2}$, which would
result in a very low radiative efficiency. Overall it seems unlikely
that self absorbed synchrotron emission produces the observed 
spectral slopes $>1/3$.

So far we have neglected Inverse Compton (IC) scattering.  For
$\epsilon_e<\epsilon_B$ the effects of IC are small, as the total
power, $P_{IC}$, emitted via IC is smaller than via synchrotron:
$P_{IC}<P_{syn}$. For $\epsilon_e>\epsilon_B$ IC becomes important as
$P_{IC}/P_{syn}=\sqrt{\epsilon_e/\epsilon_B}$ \citep{SNP96}. This
additional cooling causes $\gamma_c$ and $\nu_c$ to decrease by
factors of $\sqrt{\epsilon_e/\epsilon_B}$ and $\epsilon_e/\epsilon_B$,
respectively. This increases the duration of the FC stage in ESs by a
factors of $\epsilon_e/\epsilon_B$ and $\sqrt{\epsilon_e/\epsilon_B}$
for ISM and stellar wind environments, respectively.
 
Our results depend on the assumption of an orderly layered structure
behind the shock: a thin un-cooled layer of width $l'_{0}$, followed
by a much wider layer of cooled electrons, of width
$\Delta'=l'_{0}\sqrt{\nu_m/\nu_c}$. Clearly, significant mixing would
homogenize the region and would lead to the ``homogeneous'' spectrum
given in SPN, without the $\nu^{11/8}$ region discussed here. A
typical electron is not expected to travel much farther than its
gyration radius, $r(\gamma)$. We obtain $r(\gamma_m)/l'_{0}=5\times
10^{-8}t_{2}^{-9/8}, 10^{-9}t_{hr}^{-5/4}$ and $5\times 10^{-8}$,
using the scalings of eqs.  (\ref{ES_ISM}) and (\ref{ES_wind}) for ESs
and eq.  (\ref{IS}) for ISs, respectively. Thus, this effect could not
cause significant mixing. Another mechanism that might cause mixing is
turbulence. The observation of a spectrum with
$F_{\nu}\propto\nu^{11/8}$ would indicate the existence of a layered
structure, and constrain the effective mixing length to $\lesssim
l'_{0}=\Delta'\sqrt{\nu_c/\nu_m}$.

\acknowledgements

This research was supported by  the US-Israel BSF.

\end{document}